\documentclass[5p,number,sort&compress]{elsarticle}
\usepackage[utf8]{inputenc}
\usepackage{adjustbox}
\usepackage{xurl}
\usepackage{subcaption}
\usepackage[linkcolor=RoyalBlue, citecolor=Aquamarine, urlcolor=cyan, colorlinks=true, pdfauthor=author]{hyperref}
\usepackage[labelfont=bf,justification=raggedright,singlelinecheck=false]{caption}
\def\Plus{\texttt{+}}
\begin{document}

\begin{frontmatter}
\title{Toward asynchronous EEG-based BCI: Detecting imagined words segments in continuous EEG signals}
\author{Tonatiuh Hernández-Del-Toro\corref{cor1}}
\ead{tonahdztoro@gmail.com}
\author{Carlos A. Reyes-García}
\ead{kargaxxi@inaoep.mx}
\author{Luis Villaseñor-Pineda}
\ead{villasen@inaoep.mx}
\address{Biosignal Processing and Medical Computing Lab. Instituto Nacional de Astrofísica, Óptica y Electrónica. México.}
\cortext[cor1]{Corresponding author}
\begin{abstract}
An asynchronous Brain--Computer Interface (BCI) based on imagined speech is a tool that allows to control an external device or to emit a message at the moment the user desires to by decoding EEG signals of imagined speech. In order to correctly implement these types of BCI, we must be able to detect from a continuous signal, when the subject starts to imagine words. In this work, five methods of feature extraction based on wavelet decomposition, empirical mode decomposition, frequency energies, fractal dimension and chaos theory features are presented to solve the task of detecting imagined words segments from continuous EEG signals as a preliminary study for a latter implementation of an asynchronous BCI based on imagined speech. These methods are tested in three datasets using four different classifiers and the higher $F_1 scores$ obtained are $0.73$, $0.79$, and $0.68$ for each dataset, respectively. This results are promising to build a system that automatizes the segmentation of imagined words segments for latter classification.
\end{abstract}
\end{frontmatter}

\section{Introduction}
A Brain--Computer Interface (BCI) is a tool that allows the use of brain signals to control an external device or to emit a message. There exist several ways of implementing a BCI: (i) by the brain signal acquisition technique, (ii) by the neuroparadigm used to control the BCI, and (iii) by
their response to trigger an order, being synchronous, or asynchronous \cite{AkmalNooh2011}. There is a wide gamma of brain signal acquisition techniques, we make use of non-clinical Electroencephalography (EEG) recording headsets that have shown to be capable to build BCI as the EMOTIV EPOC\Plus {} \cite{Williams2020} and the BrainVision \cite{Krepki2007} due to its non-invasiveness, fast setup, wearability, portability and low cost in comparison with clinical EEG devices and other techniques (functional Magnetic Resonance Imaging, functional Near-Infrared Spectroscopy). Common neuro--paradigms used focus mainly on Evoked Potentials, Motor Imagery, and more recently the Imagined Speech, where the Imagined Speech is defined as ``pronouncing internally to oneself a syllable, word or phrase without producing any sound or moving any articulation"\cite{Brigham2010}.

An asynchronous BCI based on imagined speech has the advantage over other BCIs  that is more direct, and is not restricted to image abstract body movements, also that it is the most natural way of communication of human beings. In order to build an asynchronous BCI that takes EEG signals as input and uses imagined speech as neuroparadigm, we must be able to detect when the user starts to imagine the word, otherwise, the BCI would only act in certain predefined windows. This problem can be solved by being able to detect the onset and ending of Imagined Words Segments (IWS) in continuous EEG signals. The detection of the onset and ending of IWS in continuous EEG signals can lead to the detection of certain words (like up, down, left, right and select) that are relevant to the BCI actions such as moving a PC pointer, or inside an iconic navigation interface with symbols to perform specific actions like commanding the driving of a wheelchair with a defined vocabulary \cite{Torres-Garcia2016, Torres-Garcia2016a}. Despite this problem has not yet been solved, some approaches have been proposed as to detect the onset of similar neuroparadigms like imagined sound production \cite{Song2014, Song2015, Song2017, Song2017a}.

In this work, five methods of feature extraction are presented to solve the task of detecting IWS from continuous EEG signals as a preliminary study for a latter implementation of an asynchronous BCI based on imagined words that controls a PC pointer. Our work experiments with three treatments of the EEG signal: (i) wavelet decomposition, (ii) empirical mode decomposition, and (iii) only noise cleaned signal. From these treated signals, five feature sets are proposed based on energies of the frequency bands, measures of chaos theory, and fractal dimensions. This work is focused on finding the best features sets to identify IWS from continuous EEG signals. Using four different classifiers, these methods are tested on three datasets that contain IWS in continuous EEG signals.

\section{Related Work}\label{sec:RelatedWork}
Although the problem of identifying IWS from continuous EEG signals has not been solved yet, there is no known baseline to whom compare metrics. This research is based on some past works that lead the way into solving this problem.

Using a Time Error Tolerance Region, in \cite{Hernandez-Del-Toro2019} was tried to identify the onset of linguistic segments in continuous EEG signals. There have been studies on movement imagery where it was tried to identify from an EEG signal, the onset of the imagined movement in asynchronous BCIs \cite{Townsend2004}. These works aim to predict the onset by analyzing the Movement Related Cortical Potentials \cite{Pereira2018, Shakeel2015}. These studies have been promising and have lead to the building of imagined movement-based BCIs. However, the speech is generated in Broca's and Wernicke's areas and there is not a direct relation between imagined speech and real speech because the imagined speech is detected over all the brain \cite{Torres-Garcia2016a}. This leads us to the use of other features for detecting imagined speech in EEG. There also have been studies where the neuroparadigm to identify is the onset of imagined sound production \cite{Song2014, Song2015, Song2017, Song2017a}. In these works, the subject imagines a high pitch sound as well as a siren-like sound, and this is used to trigger an action in an asynchronous BCI. In the study of imagined speech, there have been several works that aim to classify words or vowels \cite{Torres-Garcia2016, Garcia-Salinas2019, Moctezuma2017, Alsaleh2017}. These works have shown promising results in this field. However, all of them deal with signals which have been already segmented by hand into imagined speech segments. It is necessary to deal with the continuous signal in order to detect the IWS. The detection of IWS can be seen as a pre-processing of signal in the sense that solving this problem will lead to an automatic segmentation of imagined speech segments in signals and with this, the classification of the word can be made. 

\section{Materials and Methods}\label{sec:Methods}
All signal cleaning, trial segmentation and feature extraction methods were implemented in Matlab. The machine learning part was implemented in Python using the Sklearn libraries.

\subsection{Datasets}
\begin{figure}[ht]
    \centering
    \includegraphics[width=0.3\textwidth]{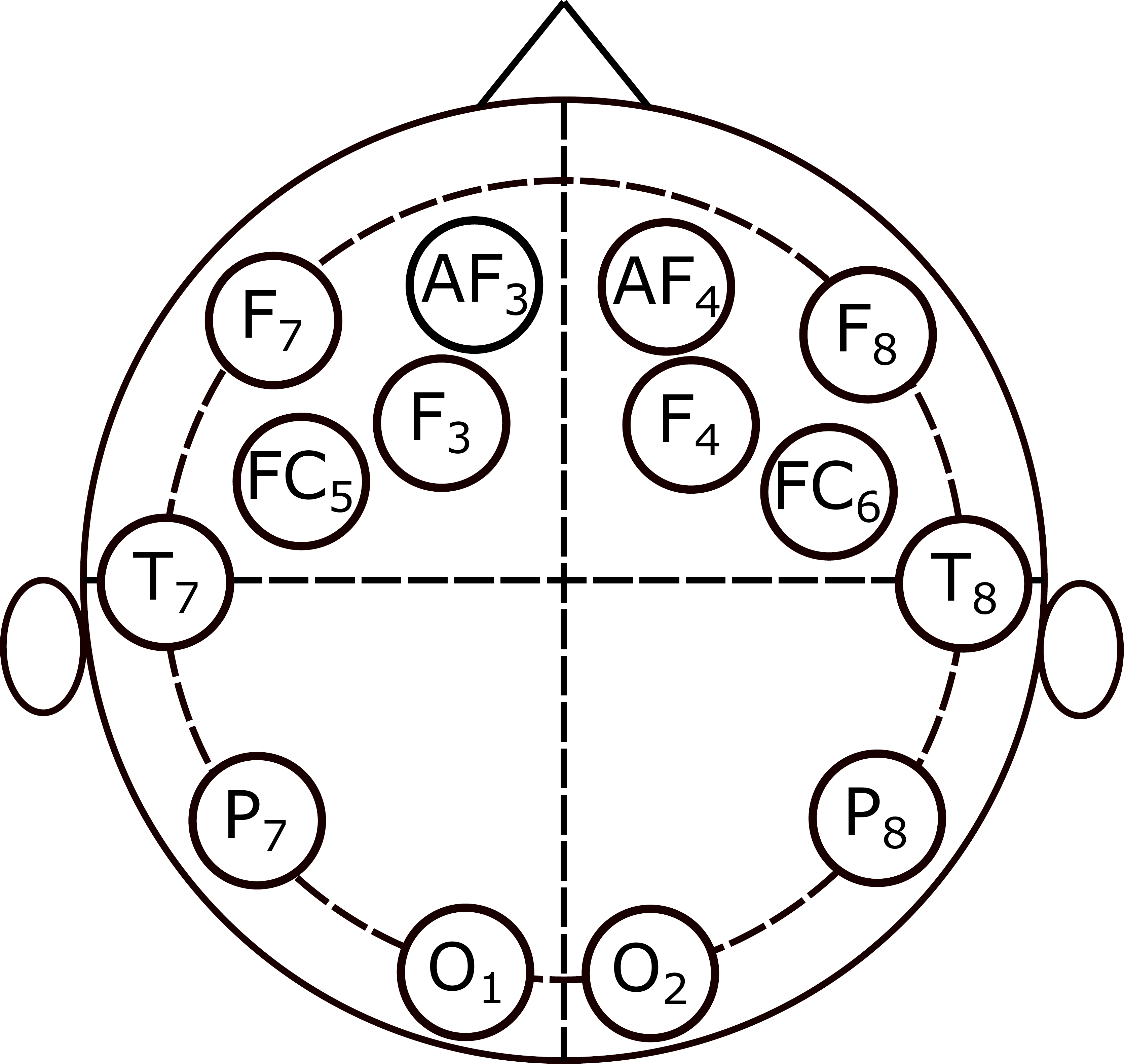}
    \caption{Placement of the 14 electrodes used in datasets 1 and 2. This same electrodes are selected in the third dataset to have the same channels.}
    \label{fig:Electrodes}
\end{figure}

Three datasets containing recordings of IWS in continuous EEG signals are used in this work. Datasets 1 and 2 were recorded using the EMOTIV EPOC\Plus {} \cite{EmotivLabs2020} headset, which has 14 channels and a sampling rate of 128~Hz, using the recording software \textit{TestBench}. Dataset 3 was recorded with the BrainVision \cite{BrainVision2020} headset which has 64 channels and a sampling rate of 500 Hz using the recording software \textit{BrainVision Professional Recorder}. Although in this third dataset the sampling rate was 500~Hz, the recordings were down-sampled to 128~Hz, and only the 14 channel used in dataset 1 and 2 were selected to have the same sample rate and channels on the three datasets, and with this, to have a better comparison of results. The reference electrodes used for recording the three datasets were P3 and P4. In \autoref{fig:Electrodes} are shown the 14 electrodes used in the three datasets.

In the three datasets, all subjects were informed about the process of recording the signals, and the privacy details in which the data can only be used for research purposes. Also, we collected the documents of informed consent that each subject read and signed before the recordings.

The first dataset consists of 27 subjects that imagine five words in Spanish which mean (up, down, left, right and select). Each subject has recordings of 5 sessions, 20 trials per session, each trial consists of an idle state followed by an imagined word, followed by another idle state. This dataset contains 100 Trials per subject and is well described in \cite{Torres-Garcia2016}, it has been also used for imagined word classification \cite{Garcia-Salinas2019}.

The second dataset consists of 27 different subjects that imagine five words in Spanish which mean (up, down, left, right and select). The main difference with respect to the first dataset is that the words that the subjects imagine, appear randomly, and were recorded in only one session. Each subject has recordings of 160 trials, each word appeared randomly 32 times. Each trial consists of an idle state given by a cross fixation, followed by the presentation of the word, then followed by the imagined word and ended by a black screen (idle state).

The third dataset consists of 20 subjects that imagine four words in Spanish which mean (up, down, left, right). Similarly to the second dataset, the words appear randomly 40 times each one in one session. Each subject has recordings of 160 Trials. Each trial consists of an idle state given by a cross fixation, followed by a picture of a house whose position indicates the word to imagine, followed by a screen showing a text that reads ``The house was", then followed by the word imagination and followed by a black screen (idle state).

\begin{figure}[ht]
    \centering
    \includegraphics[width = 0.45\textwidth]{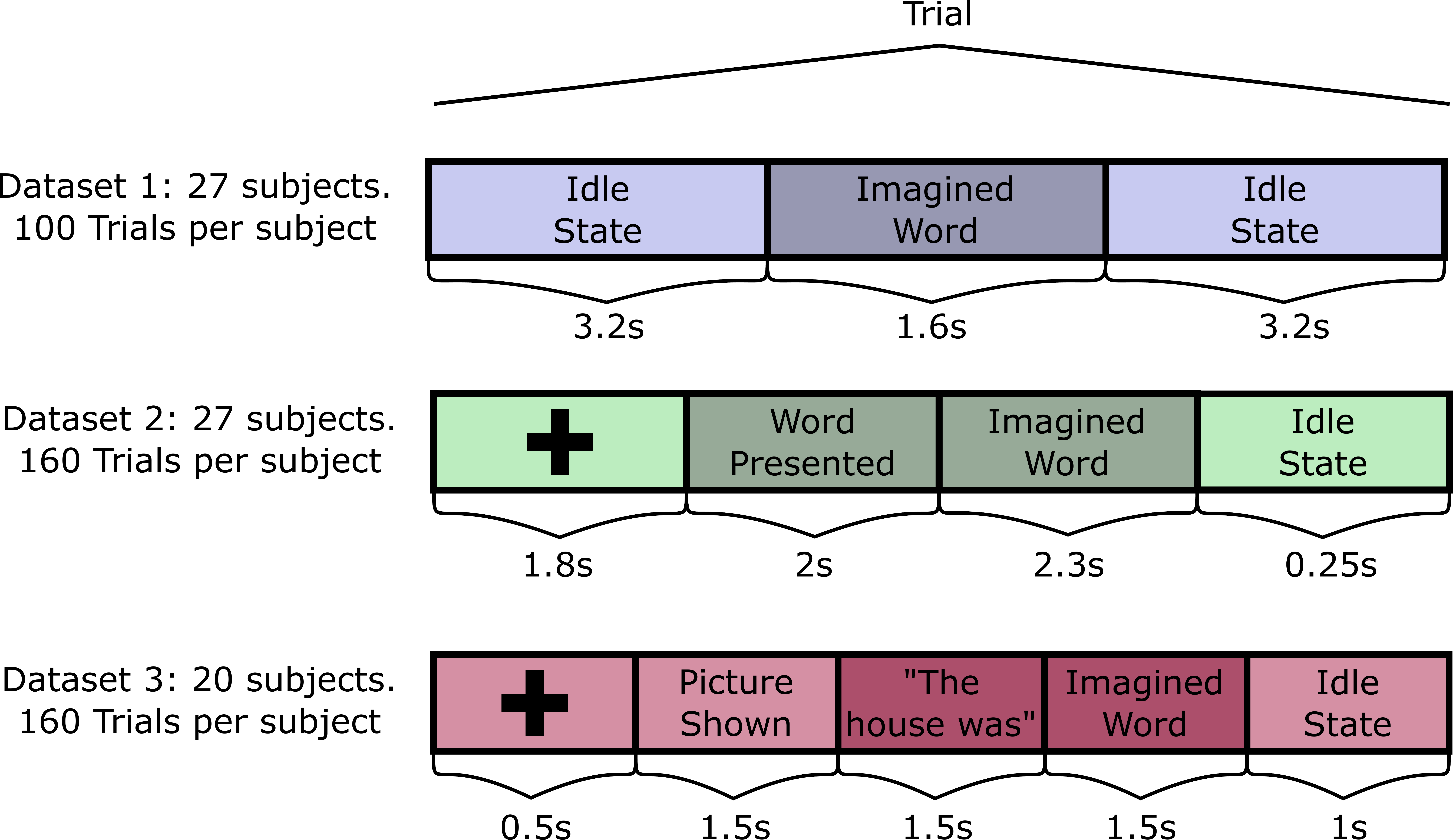}
    \caption{Protocols of recording trials in each of the three datasets, and the duration of each stage. The cross means that the subject must stare at the screen paying attention.}
    \label{fig:Protocol}
\end{figure}

The protocols of recording trials in each dataset can be seen in \autoref{fig:Protocol}, as well as the average time of each stage in the train of stimuli.

All three datasets were recorded from male and female students between ages from 20 to 30 years old. In \autoref{tab:Datasets} is summarized the description of each dataset.

\begin{table}[ht]
\caption{Description of the differences in the 3 datasets}
\centering
\begin{adjustbox}{width=0.45\textwidth}
\begin{tabular}{ccccc}
\hline
    Dataset & No. Subjects & No. of IW & Trials per word & No. Trials   \\ \hline
    1 & 27 & 5 & 20 & 100  \\  
    2 & 27 & 5 & 32 & 160  \\  
    3 & 20 & 4 & 40 & 160  \\  \hline
\end{tabular}
\end{adjustbox}
\label{tab:Datasets}
\end{table}

\begin{figure}[ht]
    \centering
    \includegraphics[width = 0.35\textwidth]{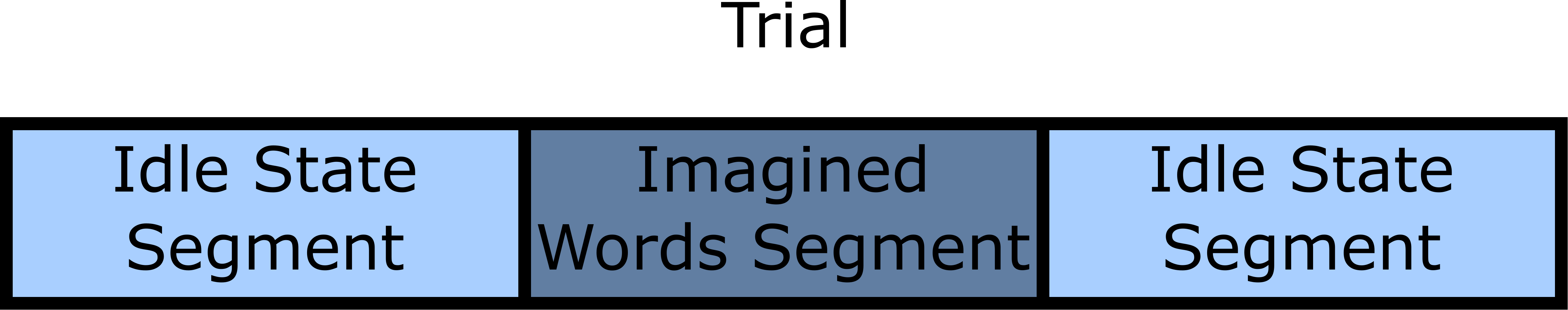}
    \caption{Example of a trial for all three datasets. It contains an ISS followed by an IWS followed by another ISS.}
    \label{fig:Trial}
\end{figure}

For dataset 1, the imagined word will be taken as an IWS while the idle states will be taken as Idle State Segments (ISS). For dataset 2, the states of cross fixation as well as the idle state at the ending will be taken as ISS. The word presented and the imagined word will be taken as IWS. For dataset 3, the states of cross fixation, the picture shown, and the idle state at the ending, will be taken as ISS. The text presented and the imagined word are taken as IWS. In this sense, for every dataset, each trial will be composed by and ISS followed by an IWS followed by another ISS as shown in \autoref{fig:Trial}.

In this scheme, all the samples prior and after the IWS are treated as ISS and the samples in between the imagined speech segment are treated as IWS, no matter which word the subject imagines, since the primary focus of the research is to identify the IWS within the trial and not to identify the word itself.

\subsection{Signal cleaning}
To reduce noise and to remove artifacts from the signals, the Common Average Reference (CAR) method is applied. This method is chosen because it has also been used in other works that deal with noise and artifacts in EEG signals \cite{Mcfarland1997, Moctezuma2017, Torres-Garcia2016, Torres-Garcia2016a, Garcia-Salinas2019, moctezuma2019a}.

\subsection{Cross validation scheme}
Imagined speech is a task that is subject dependent done and thus, using a global classifier would lead to poor performance. This property of imagined speech being subject dependent has also been used in a manner in which imagined words states and rest states of each subject are used as a biometric system for subject identification \cite{Moctezuma2018, Moctezuma2020, moctezuma2019a}. Due to this behavior, in this work, for the three datasets, an individual classifier was trained for each subject. The set of all trials of each subject was divided into a training trials set, and a test trials set, taking 75\% of trials for training the models and 25\% of trials for testing the models. This procedure was repeated 4 times and all results were averaged to obtain a 4-fold validation scheme.

\subsection{Trial Segmentation}
The segmentation  for each trial is made with a window of $0.5$ seconds that is moving with an overlap of $0.1$ seconds. This window will output a signal instance. Also, there will be two types of methods to segment the trials, these types will depend if the trial belongs to the training set or to the testing set. For the signal instances extracted either for the training set or the testing set, each signal instance will be an $64 \times 14$ matrix where 64 denotes the number of samples and 14 denotes the number of channels.

\subsubsection{Training set trial segmentation}
\begin{figure}[ht]
    \centering
    \includegraphics[width = 0.35\textwidth]{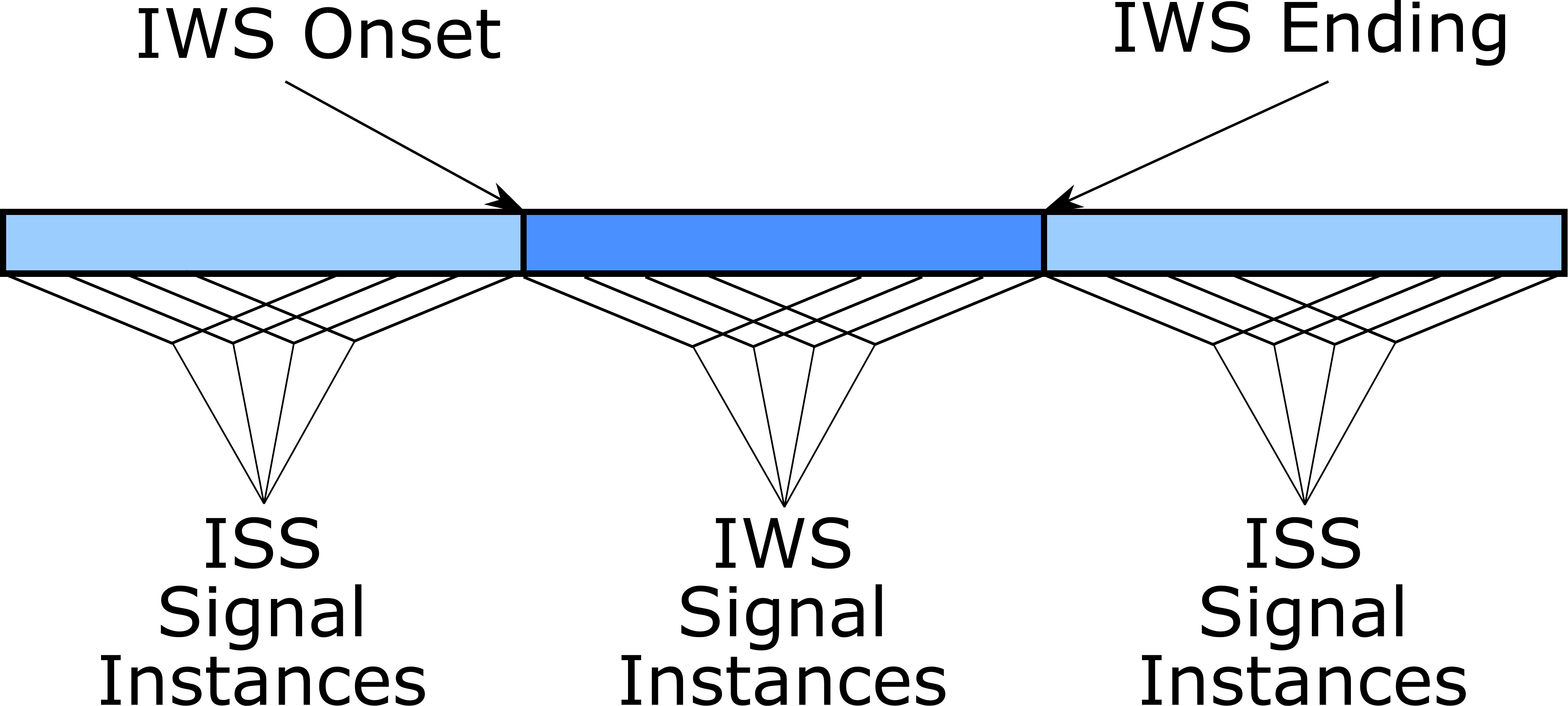}
    \caption{Segmentation of signal when the trial belongs to the training set. The signal instances of ISS are taken from two parts: (i) moving a window from the start of the trial to the onset of the IWS and (ii) moving a window from the ending of the IWS to the ending of the trial. The signal instances of IWS are obtained moving a window from the start of the IWS to the end of the IWS. In this sense, each signal instance contains only samples that belong to their class and there is no signal instance with samples of ISS and IWS simultaneously.}
    \label{fig:TrainSeg}
\end{figure}

If the trial belongs to the training set, we know \textit{a priori} the actual onset and ending of the IWS, this markers will help us to determine the number of signal instances of ISS and the number of signal instances of IWS. From the beginning of the trial, a window of $0.5$ seconds extracts a signal instance of ISS, then the window is moved by $0.1$ seconds and another signal instance of ISS is extracted, this is repeated until the end of the window reaches the onset of the IWS. Then the window of $0.5$ seconds now begins from the onset of the IWS, and from this window, a signal instance of IWS is extracted, again the window is moved $0.1$ seconds and another signal instance of IWS is extracted until the end of the window reaches the ending of the IWS. Then, the window now starts from the ending of the IWS, and from this, a signal instance of ISS is extracted, then the window is moved $0.1$ seconds and another signal instance of ISS is extracted until the end of the moving window reaches the end of the trial. This process is shown in \autoref{fig:TrainSeg}

\subsubsection{Testing set trial segmentation}
\begin{figure}[ht]
    \centering
    \includegraphics[width = 0.35\textwidth]{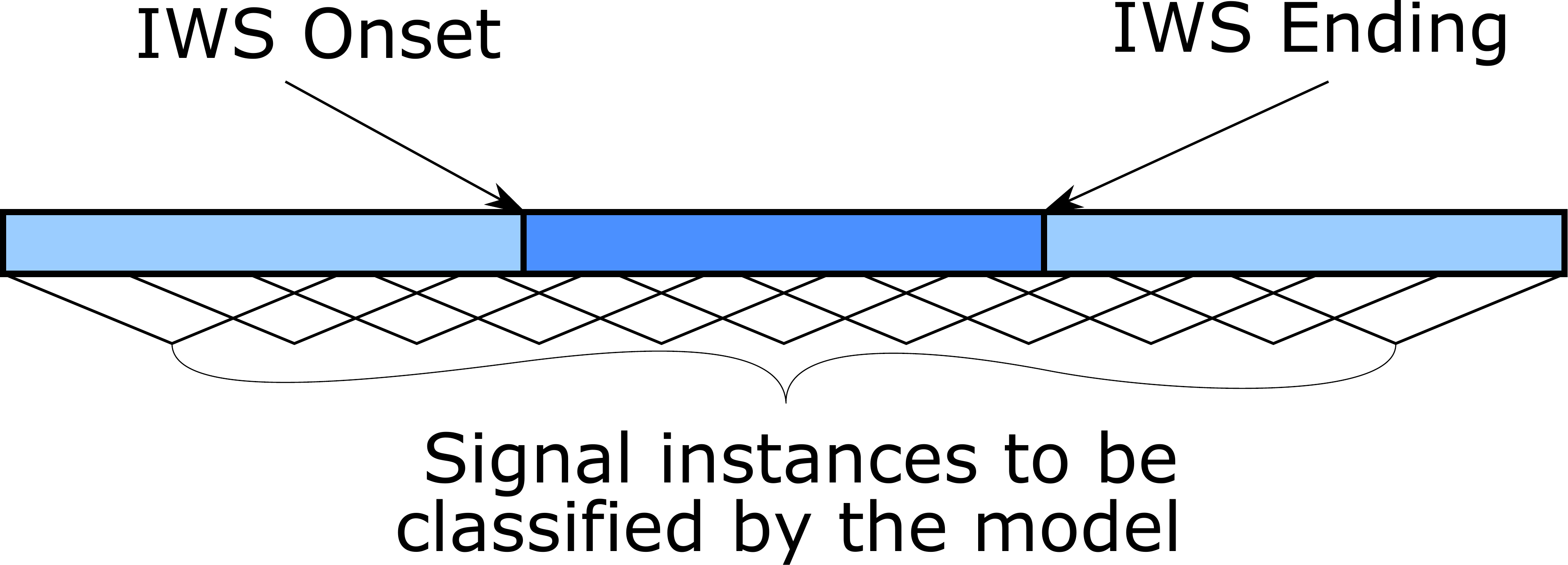}
    \caption{Segmentation of the signal when the trial belongs to the test set. In this scheme, as there is no label for the signal instances, a window will be moving from the start of the trial to the end of the trial and all signal instances will be taken no matter if they have mixed samples of ISS, and IWS simultaneously.}
    \label{fig:TestSeg}
\end{figure}

If the trial belongs to the testing set, although we know the actual onsets and ending markers, we are not going to use them, and the extraction of signal instances will be continuous over all the trial. From the beginning of the trial, a window of $0.5$ seconds will extract a signal instance and will be moved $0.1$ seconds each time until it reaches the end of trial yielding to a set of signal instances to be classified after by the model. In this scheme, some signal instances will have samples of ISS, and samples of IWS simultaneously as illustrated in \autoref{fig:TestSeg}

\subsection{Feature Extraction}
To extract features from the signal instances, first the signal is decomposed and then, the features are extracted. We decompose signals in three ways: (i) Discrete wavelet transform, (ii) Empirical mode decomposition, and (iii) Only cleaned signal. The features used in this work contain information about the frequency of signals because we know that many of the information of EEG is encoded within the frequency domain, and also the features used contain information about fractal dimensions, and chaos theory. This is due to the natural chaotic and nonlinear behavior of EEG signals.

\subsubsection{Discrete Wavelet Transform}
The discrete wavelet transform (DWT) \cite{Didiot2010} of a signal is a method of decomposing a signal into $j$ sets called detail coefficients $w_j (r)$ and an approximation coefficient $a (r)$. This coefficients give information about how each frequency band varies in time. For a signal $X(t)$, the DWT is defined as:
\begin{equation}
    DWT(n,2^j) = \sum\limits_{t=0}^{m} X(t) \Psi_{2^j}^*(t-n)
\end{equation}
with $\Psi_{2^j}(n) = \frac{1}{\sqrt{2^j}} \Psi(\frac{n}{2^j})$, where $\Psi$ is the mother wavelet and $DWT(n,2^j)$ is the discrete wavelet transform at points $n$ and $2^j$. One advantage of this method is that $DWT$ can be implemented recursively with low pass and high pass filters until a desired level $j$ of decomposition is reached. With the DWT applied over a signal using a mother wavelet $\Psi$ and a number of decomposition levels $j$ we can get a set of detail coefficients $w_j (r)$ and an approximation coefficient $a (r)$. In order to simplify the notation, the detail coefficients and the approximation coefficients are denoted as $w_j (r)$ of the $j^{th}$ decomposition level, and the approximation coefficient will correspond to the level $j+1$. In this work, 4 decomposition levels were calculated, and the approximation level was also used for feature extraction. With this, we obtain 5 sets of coefficients per channel signal. The wavelet mother used was \textit{bior2.2} because it has shown desirable results in previous works that deal with imagined speech \cite{Moctezuma2017, moctezuma2019a}.

\subsubsection{Empirical Mode Decomposition}
The Empirical Mode Decomposition (EMD) is a process where a signal is decomposed into a finite set of components called Intrinsic Mode Functions (IMF) \cite{Moctezuma2019}. The IMFs of a signal $X(t)$ are obtained through the following process:

\begin{enumerate}
    \item Calculate lower envelope $l(t)$ and upper envelope $u(t)$ of $X(t)$.
    \item Calculate the mean envelope $m(t) = \frac{u(t) - l(t)}{2}$
    \item Extract the mean from the signal to obtain $h(t) = X(t) - m(t)$
    \item Verify if $h(t)$ fulfills the two conditions of an IMF:
        \begin{itemize}
            \item The number of extrema and of zero crossings must be either equal or differ at most by one.
            \item At any point, the mean value of the envelope defined by the local maxima and the envelope of the local minima is zero.
        \end{itemize}
        If these two conditions are fulfilled, it means that $h(t)$ is an IMF. Then $h(t)$ is extracted from $X(t)$ and the process is repeated for the next IMF with the resulting signal $X'(t) = X(t) - h(t)$. Otherwise, If $h(t)$ does not fulfills the two conditions of an IMF, then the process is repeated with $h(t)$ until it is an IMF.
\end{enumerate}

Depending on the information and noise in the signal, some signals will have more IMFs than others, in order to have a constant set of IMFs from all signals, we will select the two signals with the smaller Minkowski distance because they are the IMFs that contain more information about the signal \cite{Boutana2010}. For an IMF $h(t)$ obtained from a signal $X(t)$, the Minkowski distance is defined as:

\begin{equation}
    D_M =  \left( \sum\limits_{t=1}^m \left| X(t) - h(t) \right|^2 \right)^{1/2}
\end{equation}

To maintain consistency with the detail coefficients obtained through DWT in the feature extraction process, the two selected IMFs will be named as $w_j(r) : \{ h_1(r), h_2(r) \}$. With the signal decomposed in sets $w_j(r)$ either with EMD or DWT, features can now be calculated.

\subsubsection{Instantaneous Energy}
The Instantaneous Energy (IE) \cite{Didiot2010} extracted from the detail coefficients obtained either through DWT or from the IMFs obtained from EMD reflects the energy distribution of each band. For a set $w_j(r)$ obtained through a decomposition process of a signal, the IE is defined as:

\begin{equation}\label{iwe}
    IE_j = \log_{10} \left( \frac{1}{m} \sum\limits_{r=1}^{m} w_j^2(r) \right)
\end{equation}

where $m$ is the number of samples in the set $w_j(r)$.

\subsubsection{Teager Energy}
The Teager Energy (TE) \cite{Didiot2010} reflects the variations in both amplitude and frequency of the signal. For a set $w_j(r)$ obtained through a decomposition process of a signal, the TE is defined as:

\begin{equation}
    TE_j = \log_{10} \left[ \frac{1}{m} \sum\limits_{r=1}^{m-1} \left| w_j^2(r) -  w_j(r-1) \cdot w_j(r+1) \right| \right]
\end{equation}

where $m$ is the number of samples in the set $w_j(r)$.

\subsubsection{Higuchi Fractal Dimension}
The Higuchi Fractal Dimension (HFD) \cite{Moctezuma2019, Higuchi1988} estimates the fractal dimension of a signal in the time domain. For a signal $X(t)$, the HFD is obtained as follows: First, from the original time series $X(t)$, calculate a new set of time series $X_k^m$ as:

\begin{equation}
    X_k^m = [ X(m), X(m+k), ... , X(m + [\frac{N-m}{k}]\cdot k) ]
\end{equation}

for $m=1,2,...,k$ and $k=1,k_{max}$. $k_{max}$ is usually defined between 6 and 16, in this work is fixed as 10. With the obtained new series, we calculate the length of the curve $L_m (k)$ of $X_k^m$ as:

\begin{equation}
    L_m (k) = \frac{1}{k} \left( \sum\limits_{i=1}^{[\frac{N-m}{k}]} \left| A(m,i,k) \right| \right) \left( \frac{N-1}{[\frac{N-m}{k}] k}  \right)
\end{equation}

where $A(m,i,k) = X(m + i k) - X(m + (i-1)k)$, and $N$ is the length of the original time series. The $L_m (k)$ is averaged over $m$ for each $k$, obtaining with this

\begin{equation}
    L(k) = \frac{1}{k} \sum\limits_{m=1}^k L_m(k)
\end{equation}

Finally, the HFD is estimated as the slope of the method of least-squares from the plot of $\ln (L(k))$ against $\ln ( 1/k )$.

\begin{equation}
    HFD = \frac{\ln (L(k))}{\ln ( 1/k )}
\end{equation}

\subsubsection{Katz Fractal Dimension}
The Katz Fractal Dimension (KFD) \cite{Katz1988} gives an estimation of the fractal dimension by analyzing the waveform of the time series. For a signal $X(t)$, the KFD is defined as:

\begin{equation}
    KFD = \frac{\log(m)}{\log(m) + \log(\frac{d}{L})}
\end{equation}

with $m$ the number of samples in the signal, $L$ the total length of the signal, this is:

\begin{equation}
    L = \sum\limits_{t=2}^{m} \sqrt{1 + ( X(t-1) - X(t))^2}
\end{equation}

And $d$ the planar distance of the waveform, defined as the distance from the first point to the farthest point in the signal

\begin{equation} 
    d = \max_{t} \{ \sqrt{(t - 1)^2 + (X(t) - X(1))^2} \}
\end{equation}

\subsubsection{Generalized Hurst Exponent (GHE)}
The Generalized Hurst Exponent (GHE) \cite{DiMatteo2003, DiMatteo2005, DiMatteo2007}, denoted as $H(q)$ is used in time series analysis and fractal analysis as a measure of the scaling properties by analyzing the $qth$-order moments of the distribution of the increments. For a time series $X(t)$, the GHE $H(q)$ can be obtained from the relation:

\begin{equation}
    K_q (\tau) \sim \left( \frac{\tau}{\nu} \right)^{q H(q)},
\end{equation}

with

\begin{equation}
    K_q(\tau) =  \frac{\langle | X(t+\tau) - X(t) |^q \rangle}{\langle | X(t) |^q \rangle},
\end{equation}

given from $X(t)$, with $t = \nu, 2\nu, ..., k\nu, T$. (Observation period $T$ and time resolution $\nu$). For $q=1$, the GHE is closely related to the original Hurst exponent, which measures how chaotic or unpredictable a time series is.

\subsection{Feature sets}\label{sec:FS}
Once we have a signal instance extracted from the trial, the features will be extracted from each channel, and those features will be concatenated in an instance vector. Features are calculated per each one of the following 5 feature sets:

\begin{enumerate}
    \item IE of the detail and approximation coefficients obtained with the DWT of each channel.
    \item TE, IE, HFD, KFD, and GHE with $q=1, 2$. All of them extracted from the IMFs obtained through EMD of each channel.
    \item GHE for $q=1, 2$ with only the noise cleaned signal for each channel.
\end{enumerate}

The protocol of recording the signals is different in each dataset, this can lead to a scenario where each of the feature set mentioned above performs differently for each subject in each dataset, in order to exploit the information provided by each feature, two more feature sets are defined

\begin{enumerate}
    \setcounter{enumi}{3}
    \item All the first three feature sets, where each instance is concatenated as (1, 2, 3).
    \item Principal Component Analysis (PCA) applied to the feature set no. 4 preserving 90\% of the information.
\end{enumerate}

\begin{figure}[ht]
    \centering
    \includegraphics[width=0.45\textwidth]{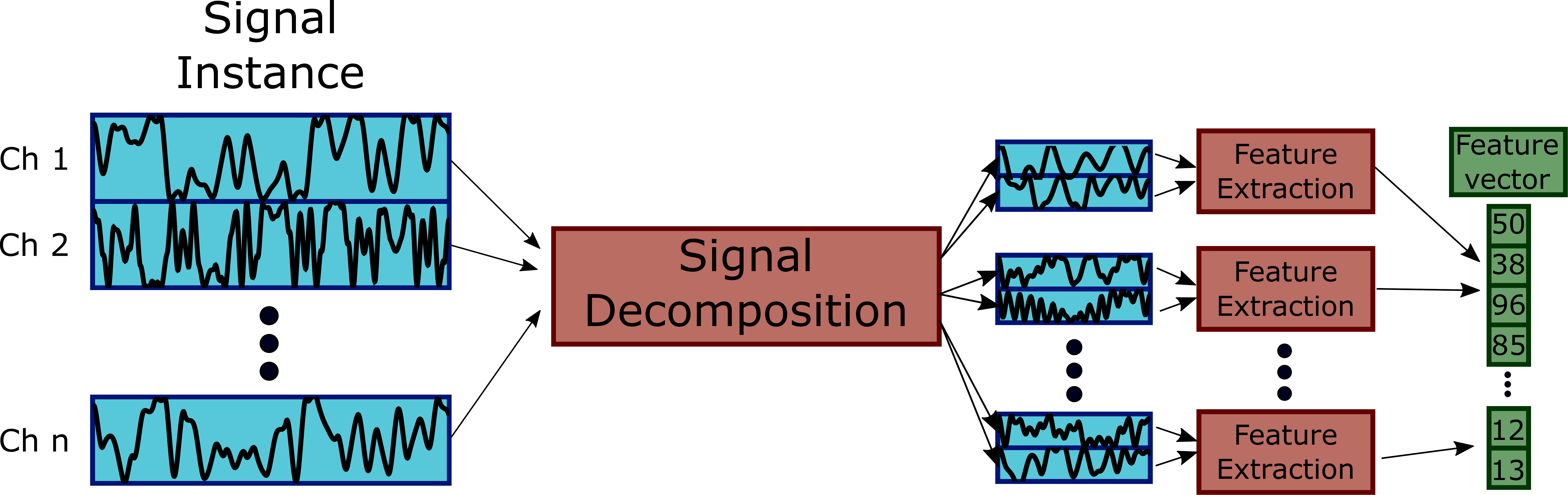}
    \caption{Procedure for extracting a feature vector from a signal instance. First each signal is decomposed for each channel, and from the decomposition, the features are calculated. The final feature vector is made by concatenation of the features obtained for each channel.}
    \label{fig:FS}
\end{figure}

The PCA is introduced to be able to exploit all the information in features but to reduce the training time of models. The feature sets based on chaos theory and fractal dimensions are selected because they have been used in studying EEG signals due to its natural non-linear a chaotic behavior \cite{Hernandez-Del-Toro2019,Vega2015,Finotello2015,Banerjee2015} and in studying epilepsy in EEG signals \cite{ Moctezuma2019, Martinez-Gonzalez2017}. The whole procedure to extract a feature vector from a signal instance is explained in \autoref{fig:FS}.

After this procedure, each signal instance of the trial will be converted into a feature vector of size $1 \times m$. The whole trial to predict will be converted into a matrix of size $n \times m$ where $n$ is the number of segments extracted from the trial and $m$ is the number of features extracted.

\subsection{Classifiers}\label{sec:Clf}
All features are normalized using the standard score $z$. The classification part was programmed in Python using sklearn libraries. Four classifiers are used in this work:

\begin{itemize}
    \item \textbf{Random Forest:} A classifier based on decision trees. For this classifier, the number of trees was 100 because was the one that performed best from the possibilities \{5, 10, 50, 100, 500\}. The number of features per tree in this model was $\sqrt{m}$ where $m$ is the number of total features.
    \item \textbf{k Nearest Neighbors:} A classifier based on the distance of the instances in the feature vector space. For this classifier, the number of neighbors was 50 because was the one that performed best from the possibilities \{5, 10, 50, 100, 500\}. The Minkowski distance is chosen to measure the distances between instances.
    \item \textbf{Support Vector Machine:} A classifier where a decision boundary is calculated. For this classifier, the default options given by sklearn were selected, using the \textit{rbf} as kernel. The regularization parameter is set to $C=1$.
    \item \textbf{Logistic Regression:} A classifier based on probability. For this classifier, the default options given by sklearn were selected, using the \textit{l2} as penalty.
\end{itemize}
Although some hyper-parameters were tuned, not many combinations were tried, since the primordial objective of this work was to test and to compare different feature extraction methods and not the classifiers.

\subsection{Window reduction and error correction}
\begin{figure}[ht]
    \centering
    \includegraphics[width = 0.45\textwidth]{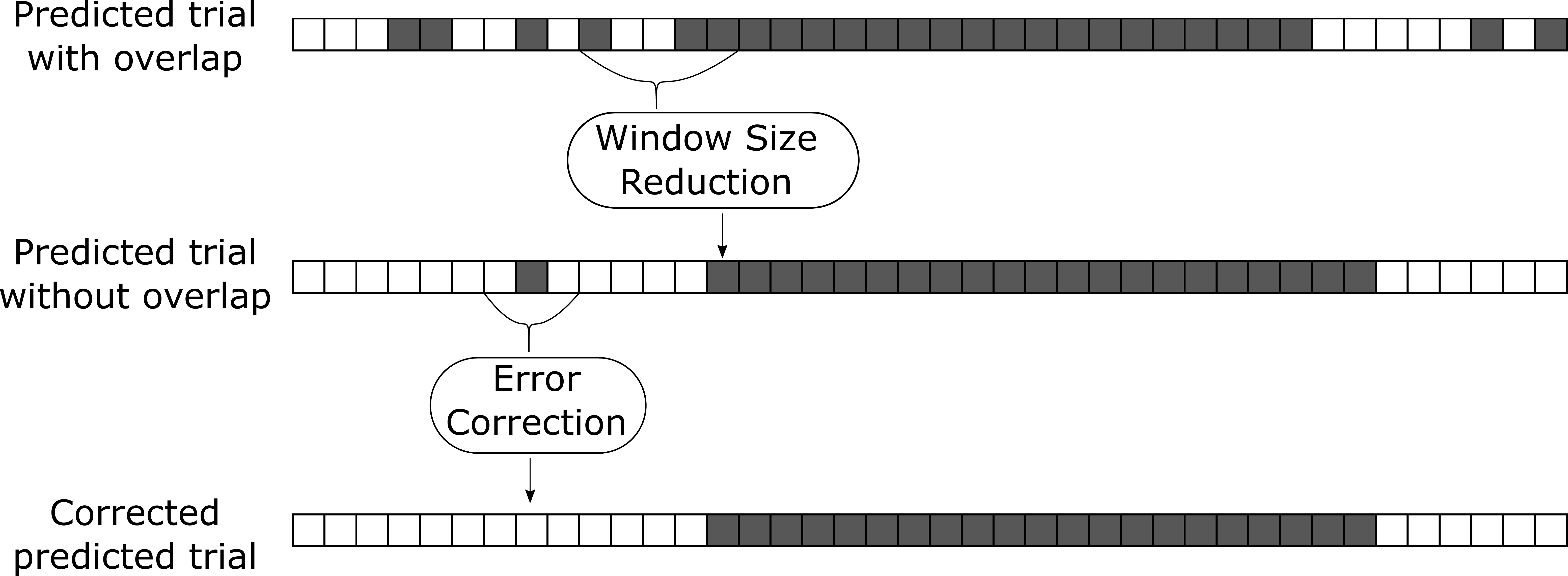}
    \caption{Procedure for converting the overlapped segments into segments of smaller window size, and procedure of error correction using first neighbors in the predicted vector to correct false positives and false negatives. Each square indicates a predicted segment. The white squares indicate that the segment belongs to an ISS, the dark squares indicate that the segment belongs to and IWS.}
    \label{fig:correction}
\end{figure}

After we have our trial classified, we obtain a vector of zeros and ones. We must remember that this vector contains information about overlapped windows. Before getting our final prediction we must turn our vector of 0.5~s sized windows and with an overlap of 0.1~s into a vector of 0.1~s sized windows with no overlap. We obtain this by making a majority vote with every 5 samples. After we do this, we also apply an error correction of first neighbors in order to delete false positives or false negatives. This whole procedure is illustrated in \autoref{fig:correction}

\subsection{Evaluation metrics}
We can compare the corrected predicted trial with the actual trial using the $F_1 score$ \cite{Tharwat2018}. The $F_1 score$ gives a trade-off of the $Precision$ and $Recall$ by penalizing if either one or another is low. It is defined as the harmonic mean of $Precision$ and $Recall$:

\begin{equation}
    F_1 score = 2 \cdot \frac{Precision \cdot Recall}{Precision + Recall}
\end{equation}

With the $F_1 score$, we can compare the corrected predicted trial with the actual trial as vectors containing clusters of $0's$ and $1's$ denoting ISS and IWS, respectively. Ideally we would like to have our predicted vectors as a cluster of $0's$ followed by a cluster of $1's$ followed by another cluster of $0's$ since that is the construction of every trial. However, this will not always happen and sometimes, the cluster of $1's$ will be moved to the left or right, or will be shrunk or enlarged. In other worse cases, we will have two clusters of $1's$ indicating that there are two IWS in the trial, which is wrong. The $F_1 score$ can take these details into account and give us a good metric to evaluate our results. With an average $F_1 score$ of 0.75, the system is capable to detect this segments of imagined words for latter classification, we would expect to have this results in all three dataset.

\section{Results}\label{sec:Results}
For each subject, and individual classifier was trained, and the predicted trials were evaluated using as metrics the $Precision$, $Recall$ $(Sensitivity)$, and $F_1 score$ with the different feature sets and classifiers.

\subsection{Feature sets 1, 2 and 3}
\begin{figure*}[ht]
  \subfloat[$F_1 score$ for dataset 1]{
	   \includegraphics[width=0.32\textwidth]{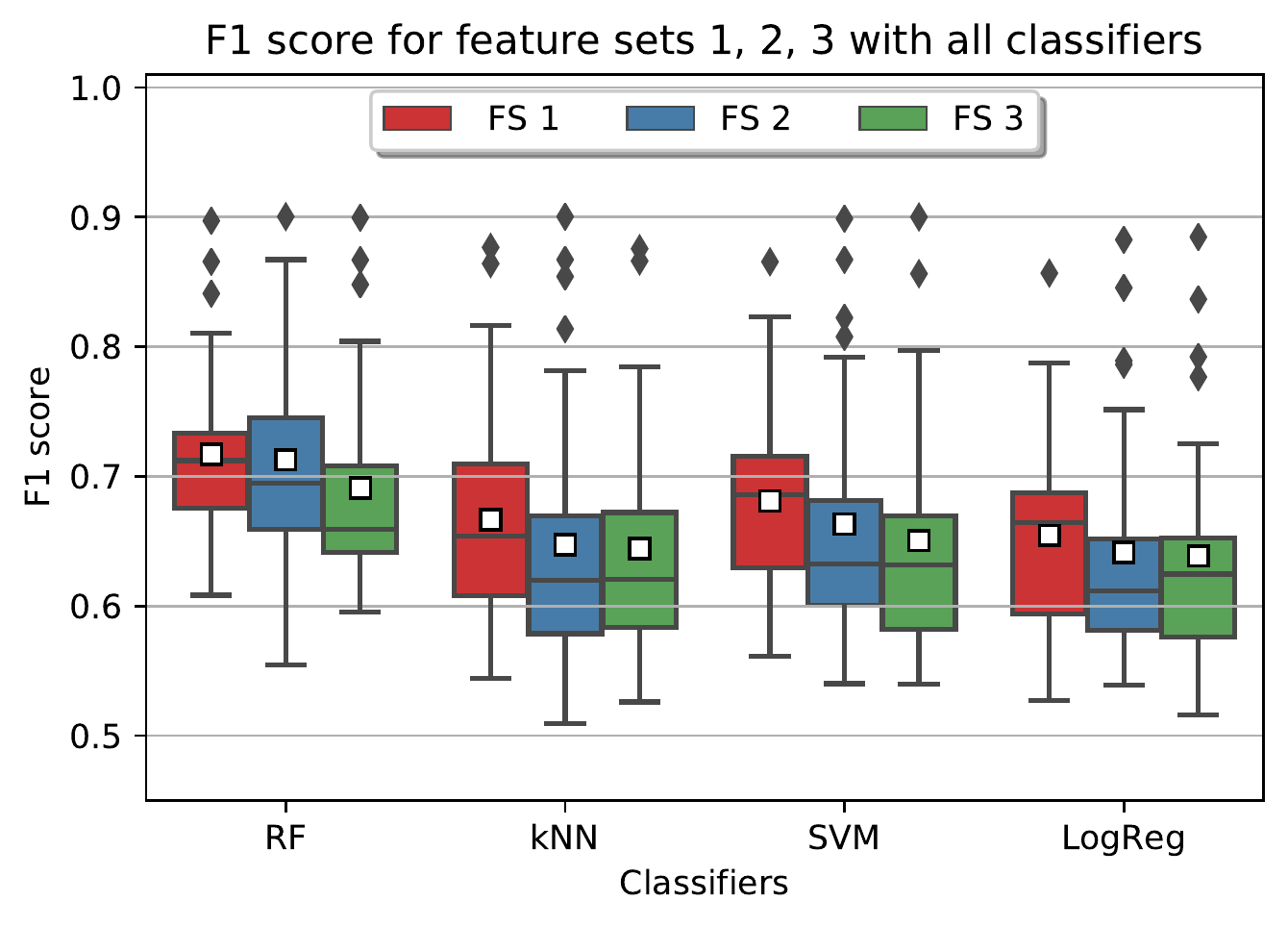}}
 \hfill 	
  \subfloat[$F_1 score$ for dataset 2]{
       \includegraphics[width=0.32\textwidth]{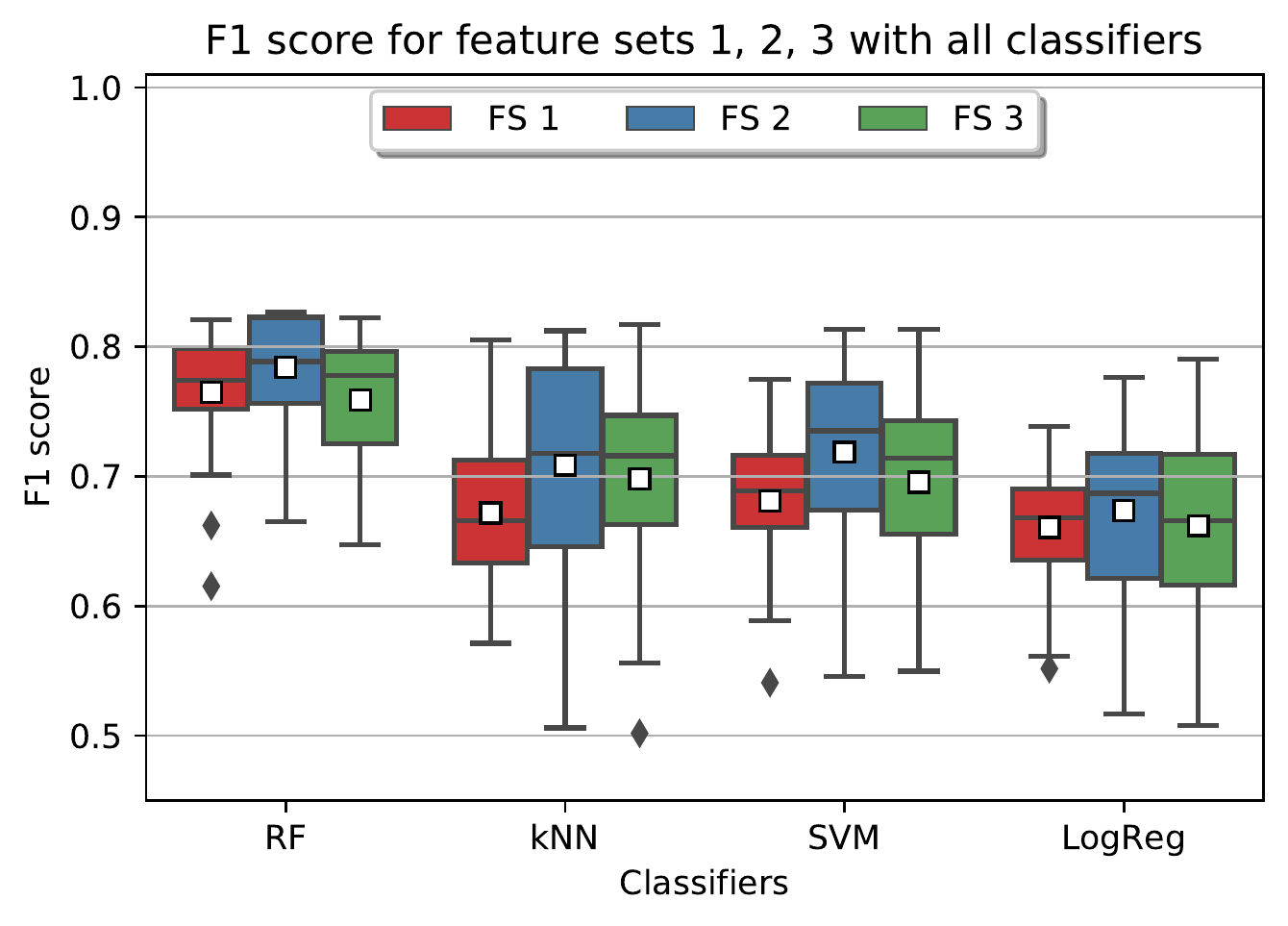}}
 \hfill	
  \subfloat[$F_1 score$ for dataset 3]{
	   \includegraphics[width=0.32\textwidth]{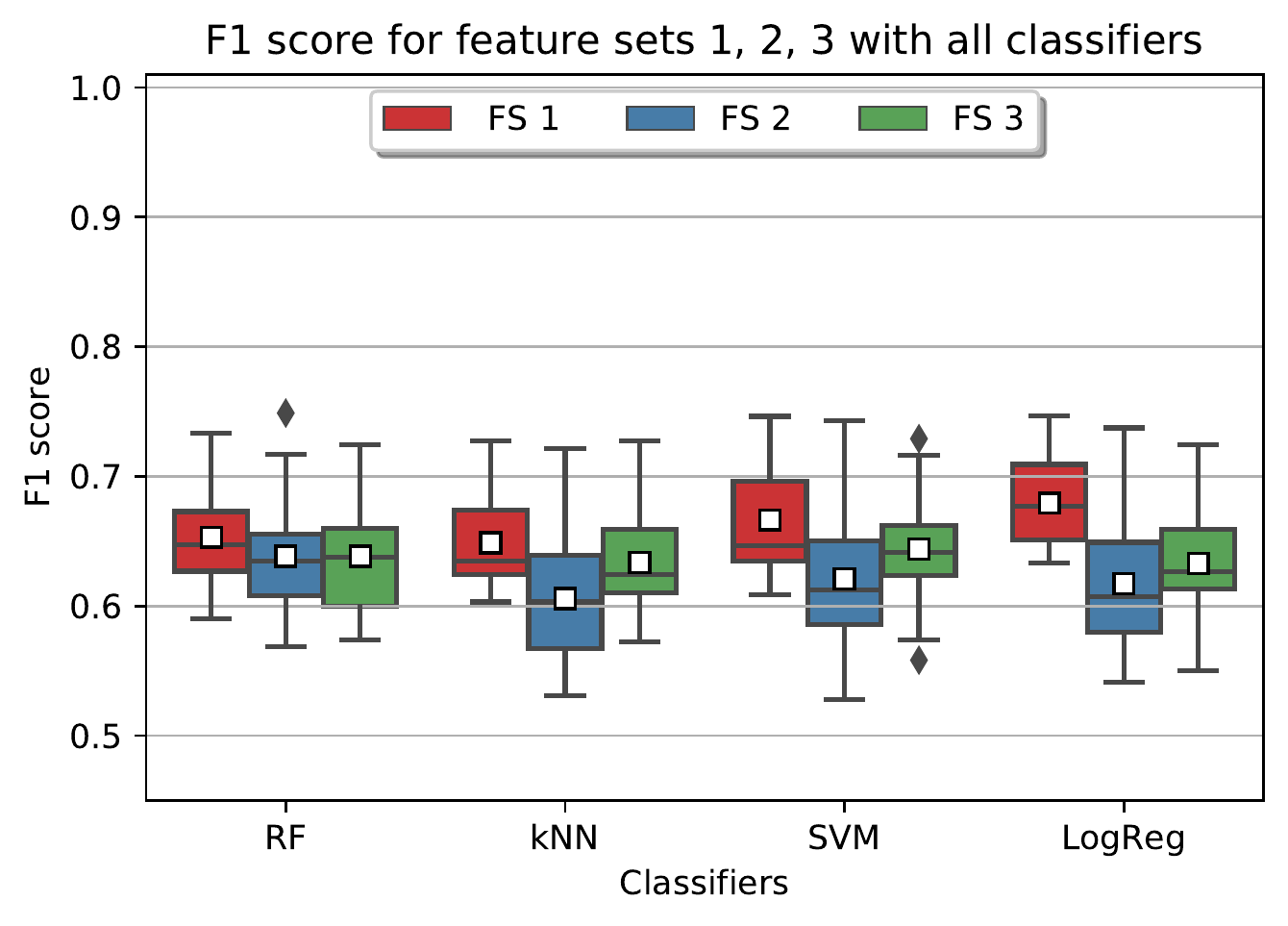}}
\caption{Results of $F_1 scores$ obtained for the three datasets using the first three feature sets described in \autoref{sec:FS} and the four classifiers described in \autoref{sec:Clf} using a box plot. The white squares inside the boxes indicate the mean obtained. The black diamonds represent subjects which result is far away from the main quarters, these elements are considered as outliers.}
\label{fig:Results1}
\end{figure*}

For the three datasets, the results for the first three feature sets are shown in \autoref{fig:Results1}. We can see that for each dataset, different feature sets perform differently. For the first dataset, the first feature set performs better in the four classifiers, although many outliers are observed on this dataset. The second dataset presents higher $F_1 scores$ when using the random forest classifier compared with the first and third dataset. And in this second dataset, the second feature set performs better in all classifiers. However, this dataset also presents outliers. In the third dataset, the $F_1 scores$ obtained are lower compared with dataset 1 and 2. Although there are not many outliers as in datasets 1 and 2. Still, there are feature sets that perform better with some subjects than others. In \autoref{tab:Results1} are summarized the best average and standard deviation results obtained, and the classifier that obtained those values.

\begin{table}[ht]
\caption{Mean and standard deviation of the best results obtained for each dataset. The value in parentheses represents the classifier and the feature set that obtained that value}
\centering
\begin{adjustbox}{width=0.45\textwidth}
\begin{tabular}{cccc}
\hline
    Dataset & $F_1 score$ & $Precision$ & $Recall$ \\ \hline
    1 & (RF, 1) 0.72 $\pm$ 0.07 & (RF, 1) 0.65 $\pm$ 0.07  & (RF, 2) 0.85 $\pm$ 0.10  \\ 
    2 & (RF, 2) 0.78 $\pm$ 0.04 & (RF, 2) 0.69 $\pm$ 0.03  & (RF, 2) 0.93 $\pm$ 0.06  \\ 
    3 & (LR, 1) 0.68 $\pm$ 0.03 & (SVM, 1) 0.62 $\pm$ 0.05  & (LR, 1) 0.85 $\pm$ 0.04  \\ \hline
\end{tabular}
\end{adjustbox}
\label{tab:Results1}
\end{table}

\begin{figure*}[ht]
  \subfloat[$F_1 score$ for dataset 1]{
	   \includegraphics[width=0.32\textwidth]{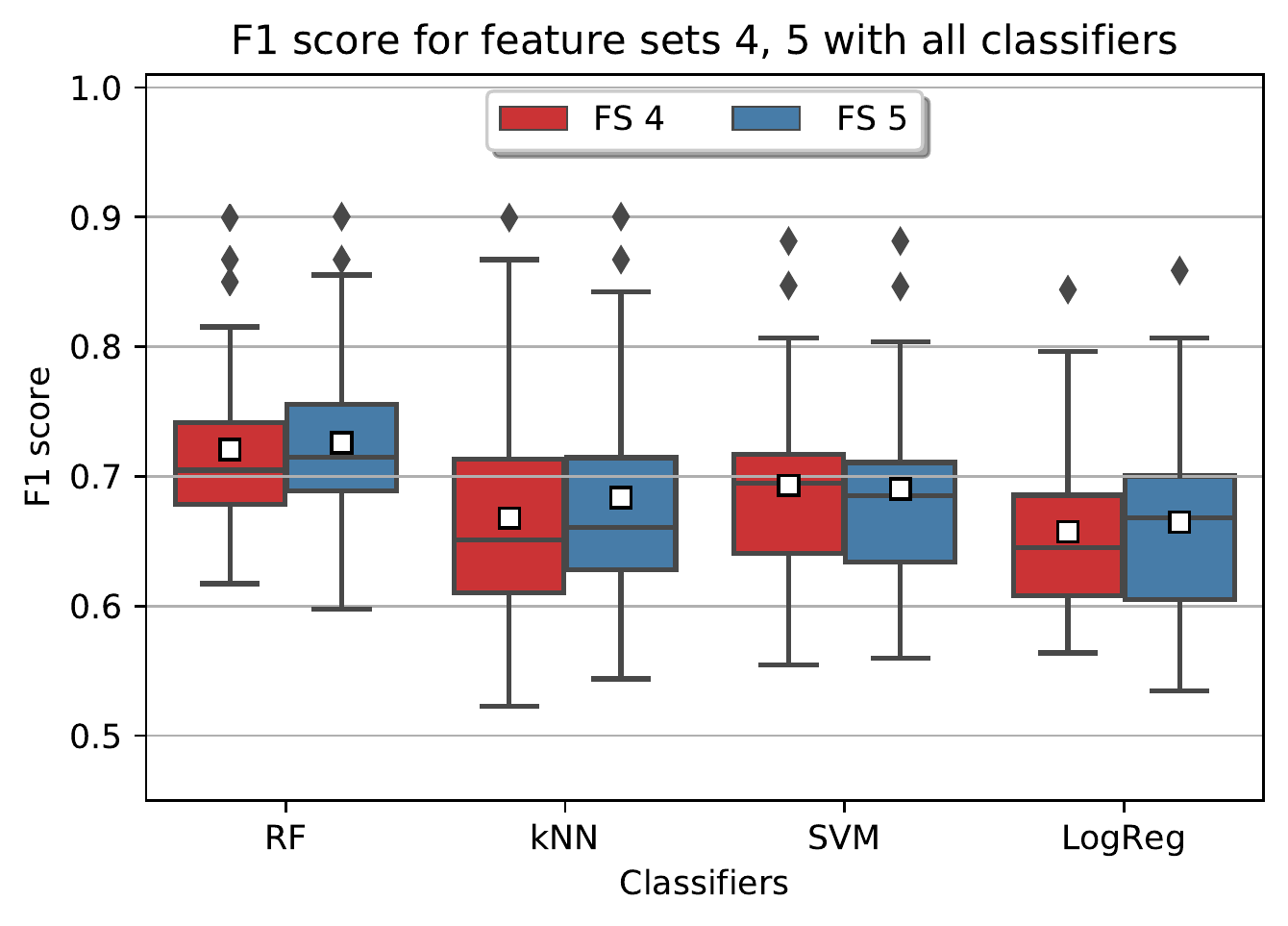}}
 \hfill 	
  \subfloat[$F_1 score$ for dataset 2]{
       \includegraphics[width=0.32\textwidth]{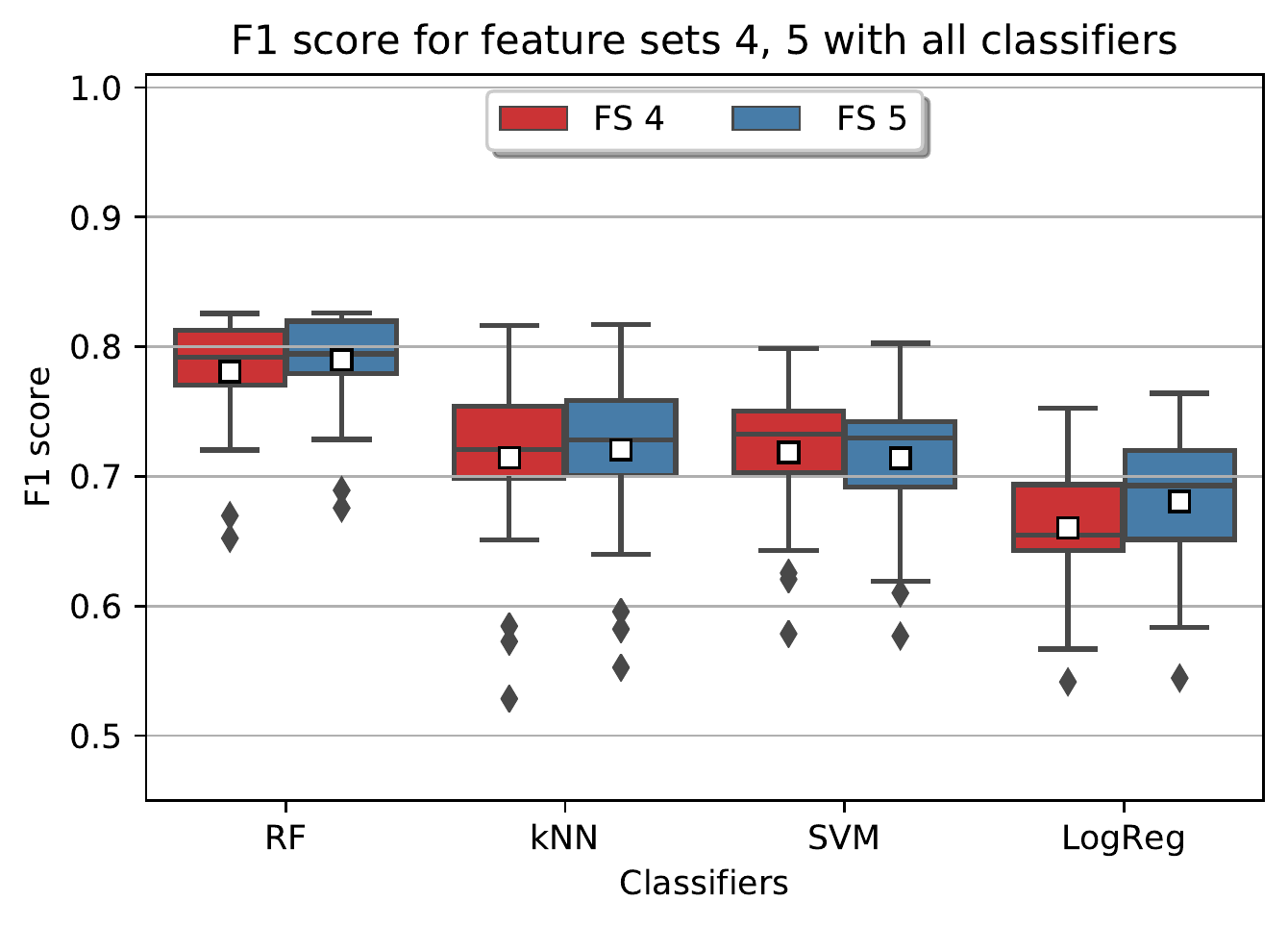}}
 \hfill	
  \subfloat[$F_1 score$ for dataset 3]{
	   \includegraphics[width=0.32\textwidth]{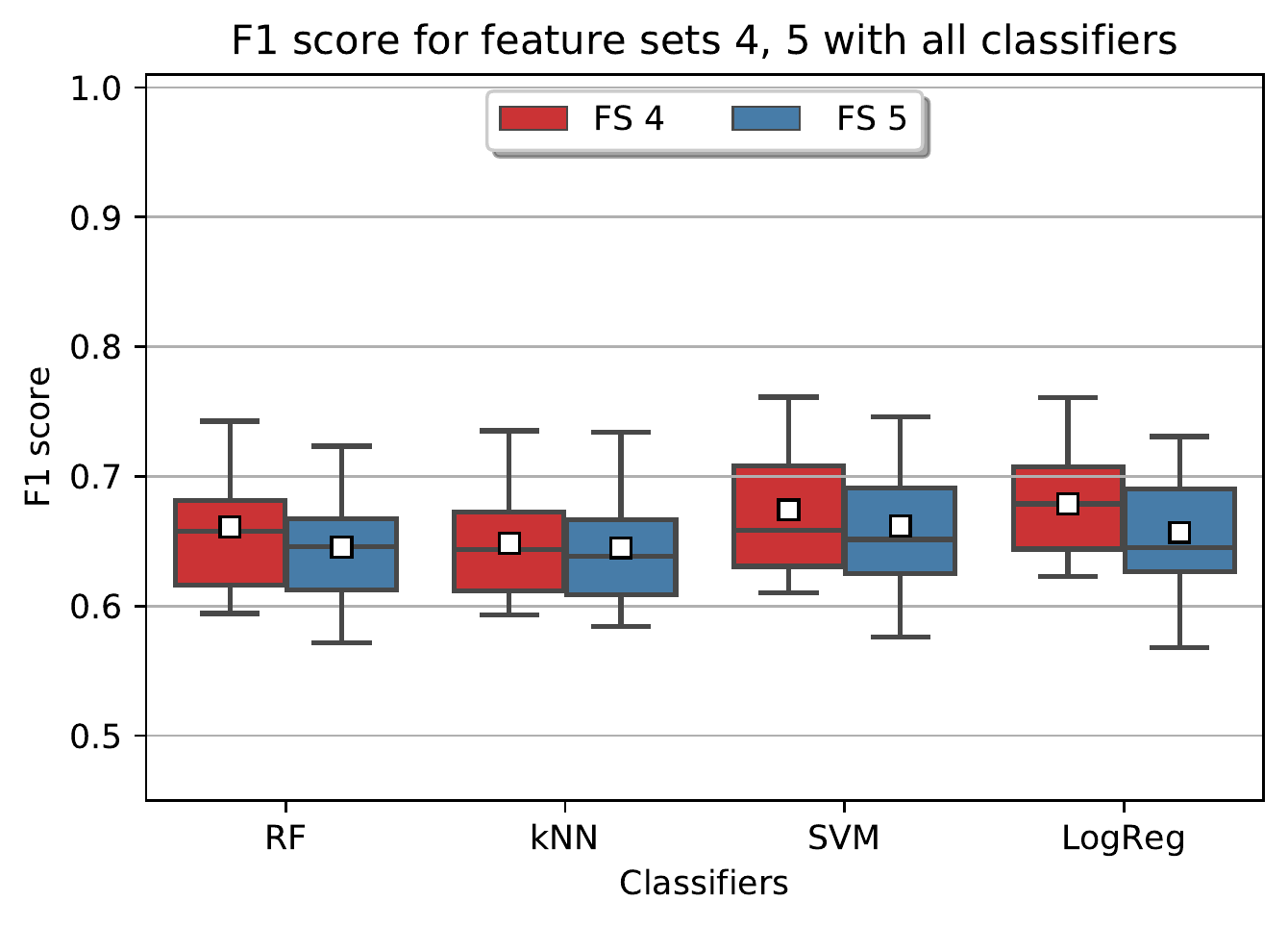}}
\caption{Results of $F_1 scores$ obtained for the three datasets using the last two feature sets described in \autoref{sec:FS} and the four classifiers described in \autoref{sec:Clf} using a box plot. The white squares inside the boxes indicate the mean obtained. The black diamonds represent subjects which result is far away from the main quarters, these elements are considered as outliers.}
\label{fig:Results2}
\end{figure*}

\subsection{Feature sets 4 and 5}
We must recall that each point in the box plot represents a subject. Thus, each box represents the data for all the subjects in the dataset. We can see that for every subject and every dataset, each feature set performs different, this is because each dataset has different recording protocols. To solve this problem and to be able to exploit the information present in all feature sets, the feature sets 4 and 5 described in \autoref{sec:FS} are used. The feature set 4 is used to exploit all the features and the feature set 5 is used to test if it is possible to get similar results with less features, thus, speeding up the training times. In \autoref{fig:Results2}, the results for the three datasets using the last two feature sets and the four classifiers are presented. We can observe that for dataset 1 not only fewer outliers appear but also the performance has increased. We can also notice that when using the feature set which has all the features plus a feature selection based on PCA, the performance is better than using all the features. For the second dataset, we can observe that contrary to the first dataset, more outliers have appeared in comparison to the first three feature sets. However, the $F_1 score$ has increased with the use of all the features, though the change is not as great as in the other datasets. The classifier that achieves the higher $F_1 score$ is the random forest. For the third dataset, no outliers appear, and we can see that using all the features raises the performance. Although the use of the PCA for feature selection reduces the features, this does not increase the performance. In \autoref{tab:Results2} is shown the best scores obtained for $F_1 score$, $Precision$, and $Recall$ with each classifier in all three datasets for this last two feature sets.

\begin{table}[ht]
\caption{Mean and standard deviation of the best results obtained for each dataset. The value in parentheses represents the classifier and the feature set that obtained that value.}
\centering
\begin{adjustbox}{width=0.45\textwidth}
\begin{tabular}{cccc}
\hline
    Dataset & $F_1 score$ & $Precision$ & $Recall$ \\ \hline
    1 & (RF, 5) 0.73 $\pm$ 0.07 & (RF, 4) 0.66 $\pm$ 0.08  & (RF, 5) 0.87 $\pm$ 0.9  \\ 
    2 & (RF, 5) 0.79 $\pm$ 0.04 & (RF, 4) 0.70 $\pm$ 0.04  & (RF, 5) 0.93 $\pm$ 0.06  \\ 
    3 & (LR, 4) 0.68 $\pm$ 0.04 & (SVM, 4) 0.65 $\pm$ 0.06  & (LR, 4) 0.83 $\pm$ 0.04  \\ \hline
\end{tabular}
\end{adjustbox}
\label{tab:Results2}
\end{table}

\subsection{Discussion}\label{sec:Discussion}
In the first dataset, using the first three feature sets, the higher average $F_1 score$ obtained was $0.72$ with the feature set 1 and the random forest classifier. Using feature set 5, this $F_1 score$ is increased to 0.73. For the second dataset, similarly to the first dataset, first we obtain a $F_1 score$ of 0.78 using the second feature set and the random forest classifier. Then this $F_1 score$ is increased to 0.79 when using the feature set 5. In third dataset, the $F_1 score$ obtained using the last two feature sets does not change in comparison with using the first three datasets, both feature sets 1 and 4 achieve an $F_1 score$ of 0.68.

\begin{figure}[ht]
    \centering
    \includegraphics[width=0.45\textwidth]{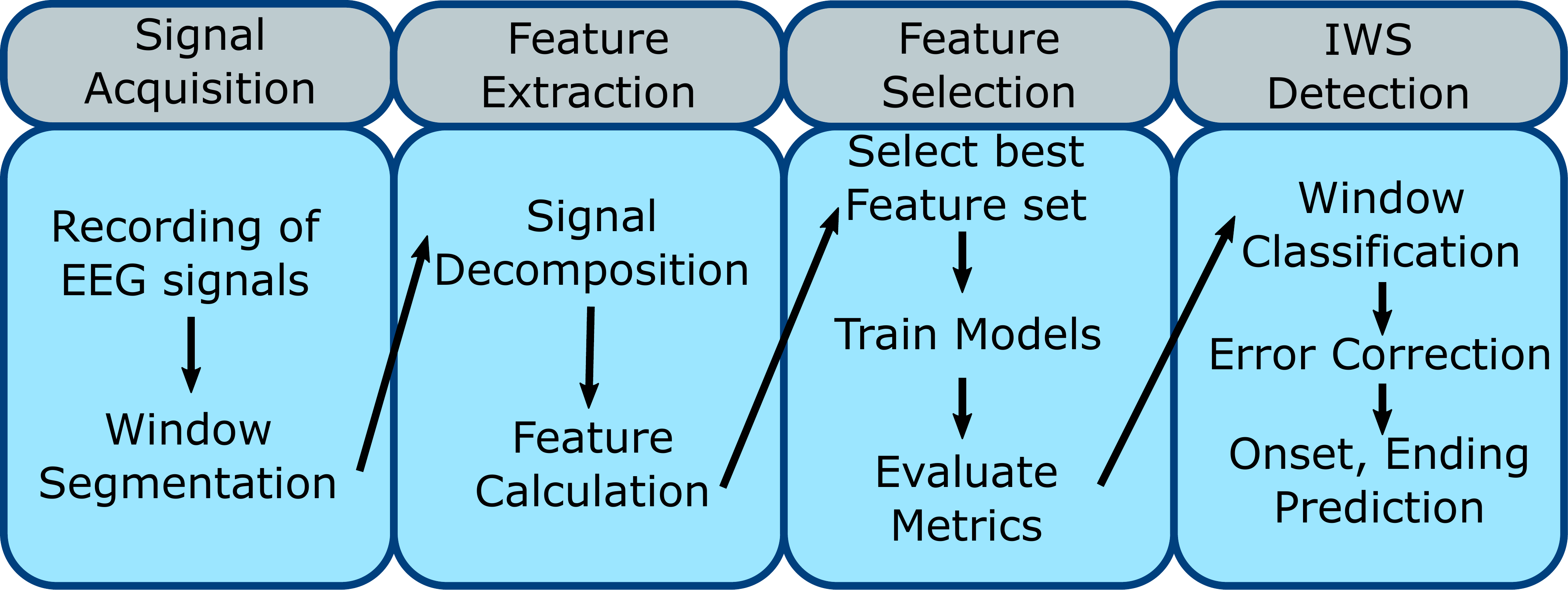}
    \caption{Proposed workflow to detect IWS exploiting the features that achieve best scores per subject.}
    \label{fig:workflow}
\end{figure}

We can see that each dataset performed differently. This behavior can be explained because each feature set performs differently with each subject. However, the use of the feature set that contains all the features keeps the scores or increases them. Although all the features analyze the frequency or fractal dimensions of EEG, the exact feature per subject can be different. This suggests the use of a model with different features per subject in order to exploit the best information given by the EEG signals in the task of detecting segments of imagined words. This feature selection can be seen as a pre-processing part of the workflow in the system that automatizes the segmentation of imagined words segments. In \autoref{fig:workflow} is shown a suggestion of this workflow.

The difference in metrics obtained in each dataset can also be explained by analyzing the protocol of recording the EEG signals. In the first dataset, the periods in between imagined words were not constant and the onset and ending of the imagined words were set by the user. In the second dataset, the timing of the idle states was controlled by the program and the size of ISS were smaller compared to the IWS, this could explain why the second dataset is the one that achieves higher scores. In the third dataset the timing was also controlled by the recording program. However, maybe the picture shown is making the subjects to imagine the word before it is needed, and this could produce a higher error. This can be corrected by recording a dataset that has different idle states before the imagined word.

\section{Conclusion}\label{sec:Conclusion}
In this work five feature sets based on DWT, EMD, Energies features, Fractal dimensions features, and chaos measure features were tested with three datasets that contain imagined words in continuous signals. The features were extracted to train four classifiers, and with them to try to identify IWS in trials that contain an ISS followed by an IWS followed by another ISS. For each subject in each dataset, different feature sets perform differently, this suggests the use of a model that selects the features that contain the most relevant information per subject. Each dataset has a different recording protocol. The dataset that achieved the higher $F_1 score$ was the one that has the imagined words more controlled. This suggests the recording of another dataset which contains idle states more controlled in the recording protocol. The results shown in this work can be used to build a system that automatizes the segmentation of IWS followed by the classification of the word itself and with this to implement a real time interactive BCI based on imagined speech that is fully asynchronous. An example of this is a BCI where the user controls a PC pointer by imagining direction words. The work presented takes part of a bigger project that aims to build a BCI based on imagined speech on EEG signals. The results shown in this paper explore the preliminary work into detecting IWS in continuous EEG signals.

\section{Future work}\label{sec:FutureWork}
As future work, we still need to implement another metrics to evaluate from the predicted trials, the onset and ending of the word in order to implement an asynchronous BCI that triggers and action when it detects the onset of an imagined word. Although feature sets 4 and 5 keep the scores obtained in feature sets 1, 2, and 3, the variation is small, this suggests that there could be other features that could exploit more the information contained in EEG signals. The average time of an EEG recording session, since the setup to the end of the recording lasted about half an hour, this duration was exhausting for the subjects as at the end many of them were tired. However, training the models with just a few samples did not perform as well as using the 75\% of trials for training. An approach that could solve this problem is the incremental learning approach in which the model is learning on the run from the signals given by the user while using the BCI, this approach would let the users to record only a few signals and start using the BCI while the BCI keeps learning from new samples. Despite the information studied in imagined speech appears in all the EEG channels, there could be some combination of channels that could keep the good scores. This suggests the study of channel selection in order to reduce the dimension of features. Another approach that can be done is the use of deep learning in the task of identifying from a continuous signal, imagined words segments. And the final goal is to perform the classification of imagined words in real time in order to integrate this identification module into a practical working BCI.

\section{Acknowledgments}
The present work was partially supported by scholarship no. 740971 granted by CONACyT, Mexico.

\section{Additional notes}
The present work is the arXiv version of the published paper \cite{Hernandez-Del-Toro2021}. The codes for the work are in the github repository \url{https://github.com/tonahdztoro/Toward_asynchronous}.

\bibliography{arXiv.bib}


\end{document}